\documentclass{article}

\usepackage{amssymb}
\usepackage{amsmath}
\usepackage{epsfig}
\usepackage{amsfonts}
\usepackage{setspace}
\def\abstract#1{\vskip 7mm
        \begin{center}{\large Abstract}\par \smallskip
                \begin{minipage}[c]{12cm}
                        \small #1
                \end{minipage}
        \end{center}
}
\def\title#1{\begin{center}{\Large\bf #1}\end{center}}
\def\author#1{\vskip 5mm \begin{center}{#1}\end{center}}
\def\address#1{\begin{center}{\it #1}\end{center}}
\makeatletter
\@ifundefined{lesssim}{}{}
\@ifundefined{gtrsim}{}{}
\def\vereq#1#2{\lower3pt\vbox{\baselineskip1.5pt \lineskip1.5pt
\ialign{$\m@th#1\hfill##\hfil$\crcr#2\crcr\sim\crcr}}}
\makeatother

\begin{document}

\title{%
A general method for solving light-like geodesics}
\author{%
  Ren-Qi Pan\footnote{E-mail:renqipan@zju.edu.cn},
  Xi He \footnote{E-mail:jin\_hexi@126.com}
   }
\address{%
  $^1$ Hangzhou Parity Microelectronics Technology Co,.Ltd., Hangzhou, Zhejiang 311611, China\\
 $^2$ Department of Physics, Hangzhou Normal University,
Hangzhou, Zhejiang 310036, China
 }

\abstract{
  A universal method to solve the differential equations of light-like geodesics is developed. The validity of this method depends on a new theorem, which is introduced for light-like geodesics in analogy to Beltrami's ``geometrical'' method for time-like geodesics. we apply the method to the Schwarzschild and Kerr spacetime as two examples. The general solutions of the light-like geodesic equations in the two spacetimes are derived straightforwadly. After setting $\theta=\pi/2$, the general light-like geodesics in Schwarzschild spacetime reduce to the same expression as that in literatures. The method developed and results obtained in this paper may be useful in modeling dynamical phenomena in strong gravitaional fields like black holes since the solutions are expressed in terms of elliptic integrals, which can be calculated effectively.
}
\begin{center}
\end{center}

\section{Introduction}

It is difficult to solve dynamical problems of particles in general relativity, even in the situation that only gravitation interaction is considered. The world lines of particles and light rays are geodesics if they are passing through field with gravity only, which is a fundamental postulate of general relativity. To obtain solutions, one needs to integrate the geodesic equations. There are two conventional and common methods for evaluating geodescis, Lagrangian equations and Halmilton-Jacobi equations.

Chandrasekhar \cite{s3} and Rauch \cite{s4} had found the solutions of time-like geodesic equations and expressed them in terms of elliptic integral. \v{C}ade\v{z} \cite{s5}-\cite{s6} and Gomboc \cite{s7} transformed the expressions of the results by Chandrasekhar and Rauch into Jacobi elliptic functions so that there is no branch ambiguity. Kara \cite{s8} discussed the significance of approximate symmetries in the geodesic equations for Schwarzschild metric. Hioe and Kuebel \cite{s9} found analytical solutions of orbit equations and classified them according to two parameters \cite{s10}. Dino Boccaletti et al. obtained the general solutions of the time-like geodesic equations in Schwarzschild and Kerr spacetime by using Beltrami's ``geometrical'' method \cite{s11}, and they derived the relationship between variables $r$ and $t$, $r$ and $\theta$, and $\theta$ and $\varphi$.

As for light-like geodesics, the solutions of geodesic equations were expressed in terms of elliptic integrals by Chandrasekhar, Rauch and Blandford. Then, \v{C}ade\v{z} and Kosti\'{c} \cite{s12} transformed their results into Jacobi elliptic functions and got simple solutions for all three types of the orbit equations in Schwarzschild spacetime. However, they had not given any solutions to the time equations.

In most previous literatures \cite{s15,s16,s17}, geodesic equations are solved  in the condition of setting $\theta=\pi/2$ at first. The general solution to geodesic equations is still a chanllenge.
In this paper, we will try to develop a general method  to find the relationship between the coordinate variables $r$ and $t$, $r$ and $\theta$, and $\theta$ and $\varphi$ of the light-like geodesics. A new method for solving light-like geodesics is developed in section 2. Then we discuss the light-like geodesics in Schwarzschild and Kerr spacetime as two examples in the section 3 and section 4, respectively, but it's worthwhile to note that the method developed here is universal and can be applied to any other spacetimes, like anti-de Sitter space and  cosmological Friedmann-Robertson-Walker metric.

\section{Differential geometrical method for geodesic equations}
\label{theorem}
In this section we develop a general method for solving light-like geodesics, following Beltrami's theorem for time-like geodesics.
It's well known that in an n-dimensional semi-Riemannian manifold $V_{n}$, the line element can be written as
\begin{equation}
ds^{2}=g_{\mu\nu}ds^{\mu}ds^{\nu}.
\end{equation}
Two invariants in this manifold can be introduced in the form
\begin{equation}
\triangle_{1}U=g^{\mu\nu}\frac{\partial U}{\partial x^{\mu}}\frac{\partial U}{\partial x^{\nu}}=g^{\mu\nu}U_{, \mu}U_{, \nu},
\end{equation}
\begin{equation}
\triangle(U,V)=g^{\mu\nu}\frac{\partial U}{\partial x^{\mu}}\frac{\partial V}{\partial x^{\nu}}=g^{\mu\nu}U_{,\mu}V_{,\nu},
\end{equation}
where $U$ and $V$ are any real functions of $x^{\mu}(\mu=1,2,\cdots n)$. The invariants $\triangle_{1}U$, $\triangle(U,V)$ are named as Belmitrami's differential parameters of the first order. The solutions of  equations $U=const.$ and $V=const.$ are two $(n-1)$-dimensional hypersurfaces in $V_{n}$. $\triangle_{1}U$ is the square length of the gradient of $U$ and it's orthogonal to the hypersurfaces $U=const.$. The two hypersufraces $U=const.$ and $V=const.$ are orthogonal, when $\triangle(U,V)=0$.

Beltrami's theorem states \cite{s11,s13,s14}: If a complete solution of equation
\begin{equation}
\triangle_{1}U=1
\end{equation}
is known, the equations of the time like geodesics are given by
\begin{equation}
\frac{\partial U}{\partial\alpha_{i}}=\beta_{i},
\end{equation}
where $\alpha_{i}$, $\beta_{i}$ are constants.

Boccalettic et al. applied Beltrami's ``geometrical'' method to time-like geodesics in Schwarzschild and Kerr spacetime and found the general solutions of their time-like geodesic equations \cite{s11}. The advantages of Beltrami's theorem turn out in their work. But Beltrami's theorem doesn't refer to light-like geodesics and how to give a gerenal solution for light-like geodesics remains as a problem.

Inspired by Beltrami's theorem, we present a new method to solve the corresponding problems of light-like geodesics. For light-like geodesics, Beltrami's theorem must be adjusted.

We find that if a complete solution of equations
\begin{equation}
\triangle_{1}U=0
\end{equation}
is known, the equations of light-like geodesics are still given by
\begin{equation}
\frac{\partial U}{\partial\alpha_{i}}=\beta_{i},
\end{equation}
where $\alpha_{i}$, $\beta_{i}$ are constants. Next we demonstrate the theorem from Lagrangian equations.

For light-like geodesics, the line element $ds=0$ and the proper time $d\tau=0$, so them can't be Lagrangian, however, we can take $\frac{1}{2}\left(\frac{d\tau}{d\lambda}\right)^{2}$ or $\frac{1}{2}\left(\frac{ds}{d\lambda}\right)^{2}$ as the Lagrangian.
In this paper, we take $\frac{1}{2}\left(\frac{ds}{d\lambda}\right)^{2}$ as the Lagrangian,
\begin{equation}
L=\frac{1}{2}\left(\frac{ds}{d\lambda}\right)^{2}=\frac{1}{2}g_{\mu\nu}\dot{x}^{\mu}\dot{x}^{\nu}.
\end{equation}

Assuming that we know n first integrals of the equations
\begin{equation}
\dot{x}^{i}=f^{i}(x^{1},x^{2},\cdots, x^{n})
\end{equation}
and substitute them into Eq.(8), the total derivatives of Lagrangian L with respect to
$x^{i}$ is
\begin{equation}
\frac{dL}{dx^{i}}=\frac{\partial L}{\partial x^{i}}+\frac{\partial L}{\partial\dot{x}^{r}}\frac{\partial\dot{x}^{r}}{\partial x^{i}}
\end{equation}
and have
\begin{equation}
\frac{d}{d\lambda}\left(\frac{\partial L}{\partial\dot{x}^{r}}\right)=\frac{\partial}{\partial x^{i}}\frac{\partial L}{\partial\dot{x}^{r}}\dot{x}^{i}.
\end{equation}
Differentiating the identity $L=\frac{1}{2}\frac{\partial L}{\partial\dot{x}^{r}}\dot{x}^{r}$, we can obtain
\begin{equation}
\frac{dL}{dx^{i}}=\frac{1}{2}\dot{x}^{r}\frac{\partial}{\partial x^{i}}\frac{\partial L}{\partial\dot{x}^{r}}+\frac{1}{2}\frac{\partial L}{\partial\dot{x}^{r}}\frac{\partial\dot{x}^{r}}{\partial x^{i}}.
\end{equation}
If we compare Eq.(12) with Eq.(10), we can find
\begin{equation}
\frac{\partial L}{\partial x^{i}}=\frac{1}{2}\dot{x}^{r}\frac{\partial}{\partial x^{i}}\frac{\partial L}{\partial\dot{x}^{r}}-\frac{1}{2}\frac{\partial L}{\partial\dot{x}^{r}}\frac{\partial\dot{x}^{r}}{\partial x^{i}}.
\end{equation}
For light rays, $dL/dx^{i}=0$, which allow us to rewrite Eq.(12) in the form
\begin{equation}
\frac{1}{2}\frac{\partial L}{\partial\dot{x}^{r}}\frac{\partial\dot{x}^{r}}{\partial x^{i}}=-\frac{1}{2}\dot{x}^{r}\frac{\partial}{\partial x^{i}}\frac{\partial L}{\partial\dot{x}^{r}}.
\end{equation}
Substituting Eq.(14)into Eq.(13), Eq.(13) becomes
\begin{equation}
\frac{\partial L}{\partial x^{i}}=\dot{x}^{r}\frac{\partial}{\partial x^{i}}\frac{\partial L}{\partial\dot{x}^{r}}.
\end{equation}
Inverting Eq.(11) and Eq.(15) into Lagrangian equations $\frac{\partial L}{\partial x^{i}}=\frac{d}{d\lambda}\left(\frac{\partial L}{\partial\dot{x}^{r}}\right)$,
we find
\begin{equation}
\dot{x}^{r}(\frac{\partial}{\partial x^{i}}\frac{\partial L}{\partial\dot{x}^{r}}-\frac{\partial}{\partial x^{r}}\frac{\partial L}{\partial\dot{x}^{i}})=0.
\end{equation}
The Eq.(16) is satisfied if there is a function $U$, which satisfies
\begin{equation}
\frac{\partial L}{\partial\dot{x}^{r}}=\frac{\partial U}{\partial x^{r}}=U_{,r},
\quad
\frac{\partial L}{\partial\dot{x}^{i}}=\frac{\partial U}{\partial x^{i}}=U_{,i}.
\end{equation}
If we set $U_{, r}=\frac{\partial L}{\partial\dot{x}^{r}}$, the Lagrangian equations is satisfied. Substituting Eq.(17) into Eq.(2), we can get
\begin{equation}
\triangle_{1}U=g^{\mu\nu}\frac{\partial L}{\partial\dot{x}^{\mu}}\frac{\partial L}{\partial\dot{x}^{\nu}}=g^{\mu\nu}g_{\mu\alpha}\dot{x}^{\alpha}g_{\nu\beta}\dot{x}^{\beta}=g_{\alpha\beta}\dot{x}^{\alpha}\dot{x}^{\beta}.
\end{equation}
For light rays, the square of line element $ds^{2}=g_{\mu\nu}\dot{x}^{\mu}\dot{x}^{\nu}=0$, thus $\triangle_{1}U=0$, Eq. (6) is satisfied.

The function $U$ contains $n-1$ constants. Differentiating Eq.(5) with respect to these constants $\alpha_{l}$ yield,
\begin{equation}
\frac{\partial\triangle_{1}U}{\partial\alpha_{l}}=2\triangle(U,\frac{\partial U}{\partial\alpha_{l}})=0,
\quad\textrm{for } l=1, 2, \cdots, n-1,
\end{equation}
which tell us that the hypersurfaces $V_{l}=\partial U/\partial \alpha_{l}=const.$ and $U=const.$ are orthogonal, so $\frac{\partial U}{\partial\alpha_{l}}=\beta_{l}(l=1, 2, \cdots, n-1)$ are the equations of the light-like geodesics. Here we demonstrate a theorem for light-like geodesics from Lagrangian equations

It's worthwhile to note that during the demonstration we don't appeal to the specific form of metrics, so this theorem is uinversal and can be applied to any spacetime in princeple. we use this theorem to solve the differential equations of light-like geodesic equations in  Schwarzschild spacetime and Kerr spacetime next two sections, but the method developed here is able to obtain light-like geodesics of any others spacetime, like  anti-de Sitter space and Friedmann-Robertson-Walker universe.
With the help of the theorem, readers will see the first integrals of light-like geodesics can be derived effectively.

\section{The light-like geodesics in the Schwarzschild spacetime}
In this section we apply the theorem generalised in last section to Schwarzschild spacetime.
Central symmetric gravitational field is described by the Schwarzschild metric. At spherical coordinates $x^{\mu}=(t,r, \theta, \varphi)$  its line element can be expressed in the form
\begin{equation}
ds^{2}=\left(1-\frac{\alpha}{r}\right)dt^{2}-\left(1-\frac{\alpha}{r}\right)^{-1}dr^{2}-r^{2}d\theta^{2}-r^{2}\sin^{2}\theta d\varphi^{2},
\end{equation}
where $\alpha=2MG$ is a constant.
Choose an affine parameter $\lambda$ as the function $U$, then replace the previous symbol $U$ with $\lambda$ and apply the Schwarzschild metric to Eq. (6) yields
\begin{equation}
\frac{r}{r-\alpha}\left(\frac{\partial\lambda}{\partial t}\right)^{2}-\frac{r-\alpha}{r}\left(\frac{\partial\lambda}{\partial r}\right)^{2}-\frac{1}{r^{2}}\left(\frac{\partial\lambda}{\partial\theta}\right)^{2}-\frac{1}{r^{2}\sin^{2}\theta}\left(\frac{\partial\lambda}{\partial\varphi}\right)^{2}=0.
\end{equation}
If we set
\begin{equation}
\lambda=A_{1}t+A_{2}\varphi+\lambda_{1}(r)+\lambda_{2}(\theta),
\end{equation}
by using the method of ``Separation of Variable'', Eq.(21) becomes
\begin{equation}
r^2\left[\frac{r}{r-\alpha}A^{2}_{1}-\frac{r-\alpha}{r}\left(\frac{d\lambda_{1}}{dr}\right)^{2}\right]=\left(\frac{d\lambda_{2}}{d\theta}\right)^{2}+\frac{A^{2}_{2}}{\sin^{2}\theta}.
\end{equation}
Since the right side is the only function of $r$ and the left side is the only function of $\theta$, we can assume both of them equal to a same constant $A^{2}_{3}$, then we have
\begin{equation}
r^2\left[\frac{r}{r-\alpha}A^{2}_{1}-\frac{r-\alpha}{r}\left(\frac{d\lambda_{1}}{dr}\right)^{2}\right]=A^{2}_{3},
\end{equation}
\begin{equation}
\left(\frac{d\lambda_{2}}{d\theta}\right)^{2}+\frac{A^{2}_{2}}{\sin^{2}\theta}=A^{2}_{3}.
\end{equation}
Solving Eq.(24) and (25), we can obtain
\begin{equation}
\lambda_{1}=\pm\int\frac{\sqrt{A^{2}_{1}r^{4}-A^{2}_{3}(r-\alpha)r}}{r(r-\alpha)}dr,
\end{equation}
\begin{equation}
\lambda_{2}=\pm\int\frac{\sqrt{A^{2}_{3}\sin^{2}\theta-A^{2}_{2}}}{\sin\theta}d\theta.
\end{equation}
We can choose the positive sign in Eqs.(26) and (27), according to the theorem that we have showed, the geodesic equations become
\begin{equation}
\frac{\partial\lambda}{\partial A_{1}}=t+\int\frac{A_{1}r^{3}dr}{(r-\alpha)\sqrt{A^{2}_{1}r^{4}-A^{2}_{3}(r-\alpha)r}}=B_{1},
\end{equation}
\begin{equation}
\frac{\partial\lambda}{\partial A_{2}}=\varphi-\int\frac{A_{2}d\theta}{\sin\theta\sqrt{A^{2}_{3}\sin^{2}\theta-A^{2}_{2}}}\equiv\varphi+\sin^{-1}[\sinh\xi \cot\theta]=B_{2},
\end{equation}
\begin{equation}
\begin{split}
\frac{\partial\lambda}{\partial A_{3}}=-\int\frac{A_{3}dr}{\sqrt{A^{2}_{1}r^{4}-A^{2}_{3}(r-\alpha)r}}+A_{3}\int\frac{\sin\theta
d\theta}{\sqrt{A^{2}_{3}\sin^{2}\theta-A^{2}_{2}}}\\*\equiv-\int\frac{A_{3}dr}{\sqrt{A^{2}_{1}r^{4}-A^{2}_{3}(r-\alpha)r}}-\sin^{-1}[\cosh\xi\cos\theta]=B_{3},
\end{split}
\end{equation}
where $B_i(i=1,2,3)$ are constants. In process of simplification we have set $A_3=A\cosh\xi$, and $A_2=A\sinh\xi$ .

To evaluate the constant $A_{1}$, $A_{2}$ and $A_{3}$, we set $\theta=\pi/2$, which already have been discussed in many relativity books. If $\theta=\pi/2$, from Eq.(25), one can find $A_{2}^{2}=A^{2}_{3}$.
If we sums Eq.(29) and (30), we can obtain
\begin{equation}
\varphi=\int\frac{A_{3}dr}{\sqrt{A^{2}_{1}r^{4}-A^{2}_{3}(r-\alpha)r}}+const.
\end{equation}
Eq.(31) can also be written in the form
\begin{equation}
\varphi=\int\frac{A_{3}dr}{r^{2}\sqrt{A^{2}_{1}-(A^{2}_{3}/r^{2})(1-\alpha/r)}}+const.
\end{equation}
We can compare Eq.(32) with corresponding canonical formula in previous literatures. In Classical Theory of Fields (by Landau and Lifshitz) \cite{s15}, the relationship between $\varphi$ and $\theta$ was deduced from Hamilton-Jacobi equation, and was written in the form(at $\theta=\pi/2$)
\begin{equation}
\varphi=\int\frac{dr}{r^{2}\sqrt{(1/\rho^{2})-(1/r^{2})(1-r_{g}/r)}}.
\end{equation}
where $\rho=\frac{cJ}{\nu}$, $r_{g}=\frac{2Mg}{c^{2}}$, $J$ is the angular momentum and $\nu$ is the frequency of the light. Note that we use $J$ and $\nu$ to denote the angular momentum and frequency for custom, instead of the notation $M$ and $\omega_{0}$, which was used in their book. We can rewrite Eq.(33) in the form
\begin{equation}
\varphi=\int\frac{Jdr}{r^{2}\sqrt{(\nu^{2}/c^{2})-(J^{2}/r^{2})(1-r_{g}/r)}}.
\end{equation}
Comparing Eq.(34) with Eq.(32), we can find that $A_{3}=J$ and $A_{1}^{2}=\nu^{2}$ in our units; a further considering about Eq.(28) leads to $A_{1}=-\nu$.

By differentiating Eq.(32), we have
\begin{equation}
\frac{d\varphi}{dr}=\frac{A_{3}}{r^{2}\sqrt{A_{1}^{2}-(A_{3}^{2}/r^{2})(r-\alpha/r)}}.
\end{equation}
If we set $u=1/r$, we can rewrite this equation in the form
\begin{equation}
\left(\frac{du}{d\varphi}\right)^{2}=\frac{1}{A_{3}^{2}}\left[A_{1}^{2}-(A_{3}^{2}u^{2})(1-\alpha u)\right].
\end{equation}
Differentiating Eq.(36) with respect to $\varphi$ and noting that $\alpha=2Mg$ gives
\begin{equation}
\frac{d^{2}u}{d\varphi^{2}}+u=3MGu^{2}.
\end{equation}
This equation is  exactly the famous equation describing the deflection of light ray, from which one can figure out the deflection angle when light rays pass through the edge of sun. Eq.(37) can also be derived by Lagrangian equations \cite{s16}. Taking $\theta=\pi/2$, we successfully derived out Eq.(37) from Eqs.(28)$\sim$(30). Although  deriving out Eq.(37) is not the main porpose of this work, it do verify the correctness of the method developed by us.

In fact, we can derive the general light-like geodesic equation from Eqs.(29)$\sim$(30) without the special condition $\theta=\pi/2$. Eqs.(29) and (30) can be re-expressed as
\begin{equation}
\cot\theta=-\frac{\sin(\varphi-B_{2})}{\sinh\xi},
\end{equation}
\begin{equation}
\sin[f(r)+B_{3}]=\frac{\cosh\xi\sin(\varphi-B_{2})}{\sqrt{\sinh^{2}\xi+\sin^{2}(\varphi-B_{2})}},
\end{equation}
where
\begin{displaymath}
f(r)=\int\frac{A_{3}dr}{\sqrt{A^{2}_{1}r^{4}-A^{2}_{3}(r-\alpha)r}}\quad\textrm{is an elliptic integral.}
\end{displaymath}

Note that we get the above two equations without restricting $\theta=\pi/2$ and these equations are the general equations of light-like geodesics in Schwarzschild spacetime. Readers can find the time-like geodesic equations in reference \cite{s11}. After comparing, one can find the light-like geodesic equations are different from the time-like geodesic equations with one term.

\section{The light-like geodesic equations in Kerr spacetime}
Next let's discuss the light-like geodesics in Kerr spacetime, which is the most common spacetime besides Schwarzschild spacetime.
The gravitational field of a rotating star is described by Kerr metric, in natural units, the line element of Kerr metric is written in the form \cite{s16}
\begin{equation}
\begin{split}
ds^{2}=(1-\frac{2Mr}{\rho^{2}})dt^{2}-\frac{\rho^{2}}{\Delta}dr^{2}-\rho^{2}d\theta^{2}-\left[r^{2}+a^{2}\sin^{2}\theta+\frac{2Mra^{2}\sin^{2}\theta}{\rho^{2}}\right]d\varphi^{2}
-\frac{4Mra\sin^{2}\theta}{\rho^{2}}dtd\varphi,
\end{split}
\end{equation}
where $\rho^{2}=r^{2}+a^{2}\cos^{2}\theta$, $\Delta=r^{2}+a^{2}-2Mr$; $M$ and $a$ are constants, which represent mass and angular momuntum per unit mass of the central star, respectively.

The contravariant form of the metric is
\begin{equation}
g^{\mu\nu}=\frac{1}{\rho^{2}}\begin{pmatrix}\Sigma^{2}/\Delta & 0 & 0 & 2Ma/\Delta\\ 0 & -\Delta & 0 & 0\\0 & 0 & -1 &0\\2Ma/\Delta & 0 & 0 & (a^{2}\sin^{2}\theta-\Delta)/\Delta \sin^{2}\theta \end{pmatrix},
\end{equation}
here, $\Sigma^{2}=r^{2}+a^{2}\sin^{2}\theta-a^{2}\Delta\sin^{2}\theta$.

Now we can apply the method that we have used to study the geodesic equations in Schwarzschild spacetime to Kerr spacetime. With the contravariant metric,
Eq.(6) can be written in the form
\begin{equation}
\frac{\Sigma^{2}}{\Delta}\left(\frac{\partial\lambda}{\partial t}\right)^{2}+\frac{4Mra}{\Delta}\left(\frac{\partial\lambda}{\partial t}\right)\left(\frac{\partial\lambda}{\partial\varphi}\right)-\Delta\left(\frac{\partial\lambda}{\partial r}\right)^{2}-\left(\frac{\partial\lambda}{\partial \theta}\right)^{2}-\frac{\Delta-a^{2}\sin^{2}\theta}{\Delta\sin^{2}\theta}\left(\frac{\partial\lambda}{\partial\varphi}\right)^{2}=0.
\end{equation}
The second term in the left side is dependent on $r$ and $\theta$, in order to avoid it, we rewrite this equation in another equivalent form
\begin{equation}
\frac{1}{\Delta}\left[(r^{2}+a^{2})\frac{\partial\lambda}{\partial t}+a\frac{\partial\lambda}{\partial \varphi}\right]^{2}-\frac{1}{\sin^{2}\theta}\left[a\sin^{2}\theta\frac{\partial\lambda}{\partial t}+\frac{\partial\lambda}{\partial\varphi}\right]^{2}-\Delta\left(\frac{\partial\lambda}{\partial r}\right)^{2}-\left(\frac{\partial\lambda}{\partial \theta}\right)^{2}=0.
\end{equation}
 As the metric is only dependent on $r$ and $\theta$, we set
 \begin{equation}
\lambda=A_{1}t+A_{2}\varphi+\lambda_{1}(r)+\lambda_{2}(\theta).
\end{equation}
Substituting this into Eq.(43), one can obtain
\begin{equation}
\frac{1}{\Delta}\left[A_{1}(r^{2}+a^{2})+A_{2}a\right]^{2}-\Delta\left(\frac{d\lambda_{1}}{dr}\right)^{2}-\left(\frac{d\lambda_{2}}{d\theta}\right)^{2}-\frac{1}{\sin^{2}\theta}\left[A_{1}a\sin^{2}+
A_{2}\right]^{2}=0,
\end{equation}
hence,\begin{equation}
\frac{1}{\Delta}\left[A_{1}(r^{2}+a^{2})+A_{2}a\right]^{2}-\Delta\left(\frac{d\lambda_{1}}{dr}\right)^{2}=\frac{1}{\sin^{2}\theta}\left[A_{1}a\sin^{2}+A_{2}\right]^{2}+\left(\frac{d\lambda_{2}}{d\theta}\right)^{2}.
\end{equation}
One could note that the left side is only dependent on $r$ and the right side is only dependent on $\theta$, so the both sides should be equal to a constant. If we set the constant as $A_{3}^{2}$, Eq.(46) becomes
\begin{equation}
\left[A_{1}(r^{2}+a^{2})+A_{2}a\right]^{2}-\Delta^{2}\left(\frac{d\lambda_{1}}{dr}\right)^{2}=A_{3}^{2}\Delta,
\end{equation}
\begin{equation}
\left[A_{1}a\sin^{2}+A_{2}\right]^{2}+\sin^{2}\theta\left(\frac{d\lambda_{2}}{d\theta}\right)^{2}=A_{3}^{2}\sin^{2}\theta.
\end{equation}
We can evaluate $\lambda_{1}$ and $\lambda_{2}$ from Eqs.(47) and (48),
\begin{equation}
\lambda_{1}=\int\frac{\sqrt{[A_{1}(r^{2}+a^{2})+A_{2}a]^{2}-A_{3}^{2}\Delta}}{\Delta}d\theta,
\end{equation}
\begin{equation}
\lambda_{2}=\int\frac{\sqrt{\sin^{2}\theta A_{3}^{2}-[A_{1}a\sin^{2}+A_{2}]^{2}}}{\sin\theta}dr.
\end{equation}
If we set
\begin{equation}
R(r)=[A_{1}(r^{2}+a^{2})+A_{2}a]^{2}-A_{3}^{2}\Delta,
\end{equation}
\begin{equation}
\Theta(\theta)=\sin^{2}\theta A_{3}^{2}-[A_{1}a\sin^{2}+A_{2}]^{2}.
\end{equation}
Eq.(44) can be written in the form
\begin{equation}
\lambda=A_{1}t+A_{2}\varphi+\int\frac{\sqrt{R(r)}}{\Delta}dr+\int\frac{\sqrt{\Theta}}{\sin\theta}d\theta.
\end{equation}
Using Eq.(7) we can obtain the light-like geodesic equations as we did in Schwarzschild metric:
\begin{equation}
\frac{\partial\lambda}{\partial A_{1}}=t+\int\frac{(r^{2}+a^{2})[A_{1}(r^{2}+a^{2})+A_{2}a]}{\Delta\sqrt{R(r)}}dr-\int\frac{(A_{1}a\sin^{2}\theta+A_{2})a\sin\theta}{\sqrt{\Theta(\theta)}}d\theta=B_{1},
\end{equation}
\begin{equation}
\frac{\partial\lambda}{\partial A_{2}}=\varphi+\int\frac{a[A_{1}(r^{2}+a^{2})+A_{2}a]}{\Delta\sqrt{R(r)}}dr-\int\frac{A_{1}a\sin^{2}\theta+A_{2}}{\sin\theta\sqrt{\Theta(\theta)}}d\theta=B_{2},
\end{equation}
\begin{equation}
\frac{\partial\lambda}{\partial A_{3}}=A_{3}\left[-\int\frac{dr}{\sqrt{R(r)}}+\int\frac{\sin\theta}{\sqrt{\Theta(\theta)}}d\theta\right]=B_{3}.
\end{equation}

Eq.(54) and (56) are the general light-like geodesic equations in Kerr spacetime.  Though we express our finial result of light-like geodesic equations in the same form as time-like geodesics, actually, the two groups of equations are different. The function $R(r)$ and $\Theta(\theta)$ of time-like geodesics \cite{s11} are, 
\begin{equation}
R(r)=[A_{1}(r^{2}+a^{2})+A_{2}a]^{2}-(r^{2}+A_{3}^{2})\Delta,
\end{equation}
\begin{equation}
\Theta(\theta)=\sin^{2}\theta (A_{3}^{2}-a^{2}\cos^{2}\theta)-[A_{1}a\sin^{2}+A_{2}]^{2}.
\end{equation}
However, the function $R(r)$ and $\Theta(\theta)$ of light-like geodesics are
\begin{equation}
R(r)=[A_{1}(r^{2}+a^{2})+A_{2}a]^{2}-A_{3}^{2}\Delta,
\end{equation}
\begin{equation}
\Theta(\theta)=\sin^{2}\theta A_{3}^{2}-[A_{1}a\sin^{2}+A_{2}]^{2}.
\end{equation}

The light-like geodesic equations are different from the time-like geodesics equations, that's because for light the line element $ds=0$ while for massive particles do not.

If we let the angular momentum unit mass $a=0$, Eqs.(54) $\sim$(56) reduce to Eqs.(28)$\sim$(30). It's consistent with our expectation as Schwarzschild metric is a limit ($a$ $\rightarrow$0) of Kerr metric.

\section{Summary}
 Time-like geodesics can be obtained through Beltrami's theorem effectively, while the theorem doesn't refer to light-like geodesics. We develop Beltrami's theorem such that it is applicable to light-like geodesics, and via the generalized theorem we obtain the general solutions of light-like geodesics in the Schwarzschild and Kerr spacetime without the common restriciton $\theta=\pi/2$. The theorem introduced by us  for light-like geodesics is similar to Beltrami's theorem on time-like geodesics. The general light-like geodesic equations are deduced in a straightforward  way, and the elliptic integrals in the expressions of these equations can be numerically calculated efficiently and accurately with Landen transformation or Carlson's algorithms. These equations may be very useful for modeling dynamical phenomena near black hole, especially for numerical calculation.
 
  As for the light-like geodesics of Schwarzschild spacetime, after setting $\theta=\pi/2$ in the general solutions, we obtain the same results as literatures, which corroborate the validity of the new method devepled in this paper. For both Schwarzschild spacetime and Kerr spacetime, the light geodesics are slightly different with the time-like geodesics. As metioned before the method developed here is universal and is applicable to any other spacetime, like anti-de Sitter space and  Friedmann-Robertson-Walker universe. The geodesics of anti-de Sitter space and  Friedmann-Robertson-Walker universe may appear somewhere else later. The general method developed here is practical for studying tracks of light rays in various spacetimes .

\begin{flushleft}
\textbf{Acknowledgements}
We would like to thank Prof. Hong-Sheng Hou and Xun-Qiang Wu who have provided their
heartful help during the completion of this work. This work is supported by National Natural Science Foundation of China (11105036,11175053,11475051)and Natural Science Foundation of Zhejiang Province(Y6110177).
\end{flushleft}


\end{document}